\documentclass{aa}
\usepackage{afterpage} 
\usepackage{balance}
\usepackage{graphicx}
\usepackage{epstopdf}
\usepackage{subfigure}
\usepackage[normalem]{ulem} 
\usepackage{float}
\usepackage[caption = false]{subfig}
\usepackage{xcolor}
\usepackage{txfonts}
\usepackage{natbib}
\usepackage{inputenc}
\usepackage{longtable, pdflscape, multirow, booktabs}

\def \inte {\emph{INTEGRAL}}

\def \maxi {\emph{MAXI}}
\def \src {\mbox{Vela~X-$1$}}

\begin{document}

   \title{Probing the stellar wind environment of \src~with \maxi}
   \titlerunning{\src~observed with \maxi}
   \author{C. Malacaria\inst{\ref{inst1}, \ref{inst2}}
          \and
          T. Mihara\inst{\ref{inst2}}
          \and
          A. Santangelo\inst{\ref{inst1}}
          \and
          K. Makishima\inst{\ref{inst2}, \ref{inst3}}
          \and
          M. Matsuoka\inst{\ref{inst2}}
          \and
          M. Morii\inst{\ref{inst2}}
          \and
          M. Sugizaki\inst{\ref{inst2}}
          }

   \institute{Institut f\"{u}r Astronomie und Astrophysik, Sand 1, 72076 T\"{u}bingen, Germany \\ \email{malacaria@astro.uni-tuebingen.de}\label{inst1}
         \and MAXI team, RIKEN, 2-1 Hirosawa, Wako, Saitama 351-0198\label{inst2}
         \and Department of Physics, The University of Tokyo 7-3-1 Hongo, Bunkyo-ku, 113-0033 Tokyo, Japan\label{inst3}}

   \date{\today}

\abstract
{Vela X-1 is one of the best-studied and most luminous accreting X-ray pulsars.
The supergiant optical companion produces a strong radiatively driven stellar wind
that is accreted onto the neutron star, producing highly variable X-ray emission.
A complex phenomenology that is due to both gravitational and radiative effects 
needs to be taken into account to reproduce orbital spectral variations.}
{We have investigated the spectral and light curve properties of the X-ray emission from \src~along the binary orbit. 
These studies allow constraining the stellar wind properties and 
its perturbations that are induced by the pulsating neutron star.}
{We took advantage of the All Sky Monitor \maxi/GSC data to analyze \src~spectra and light curves.
By studying the orbital profiles in the $4-10$ and $10-20\,$keV energy bands, 
we extracted a sample of orbital light curves (${\sim}15\%$ of the total) showing a dip 
around the inferior conjunction, that is, a double-peaked shape.
We analyzed orbital phase-averaged and phase-resolved spectra of both the double-peaked and the standard sample.}
{The dip in the double-peaked sample needs $N_{\rm H}\sim2\times10^{24}\,$cm$^{-2}$ 
to be explained by absorption alone, which is not observed in our analysis.
We show that Thomson scattering from an extended and ionized accretion wake 
can contribute to the observed dip.
Fit by a cutoff power-law model, the two analyzed samples show orbital modulation 
of the photon index that hardens by ${\sim}0.3$ around the inferior conjunction, 
compared to earlier and later phases.
This indicates a possible inadequacy of this model.
In contrast, including a partial covering component at certain orbital phase bins 
allows a constant photon index along the orbital phases, 
indicating a highly inhomogeneous environment whose column density 
has a local peak around the inferior conjunction.
We discuss our results in the framework of possible scenarios.}
{}


   \maketitle
%

\section{Introduction}
\object{Vela~X-1} is the prototype  of the class of wind-fed accreting X-ray binaries.
The source has been discovered in $1967$ \citep{Chodil+67} 
and is located at a distance of ${\sim}2\,$kpc \citep{Nagase89}. 
It is an eclipsing supergiant High Mass X-ray Binary (HMXB),
consisting of a neutron star (NS) with mass M$_{\rm NS}{\sim}1.77\,$M$_{\odot}$ \citep{Rawls+2011}
and a B0.5Ib optical companion, HD $77581$, with mass M$_{\rm B}{\sim}23\,$M$_{\odot}$ 
and radius R$_{\rm B}{\sim}30\,$R$_{\odot}$ \citep{vanKerkwijk+95}.

\begin{figure}[t!]
\includegraphics[width=\hsize]{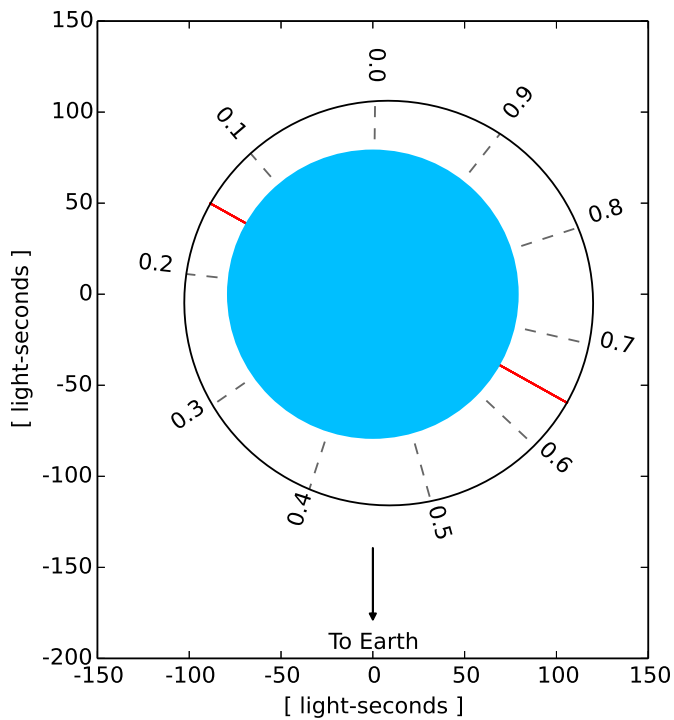}
\caption{Orbital sketch of the binary system \src~as observed from Earth. 
	The blue disk centered at ($0,0$) is the optical companion HD$\,77581$. 
	The black solid line is the NS elliptical orbit. 
	The gray dashed rays and the respective numbers indicate the orbital phases. 
	Phase zero is taken at the mid-eclipse time. 
	The red straight solid line crossing the ellipse represents the major axis, 
	pointing the periastron and the apastron at the intersections with the NS orbit.
	}
\label{fig:sketch}
\end{figure}

A sketch of the binary system is shown in Fig.~\ref{fig:sketch}.
The binary orbit is almost circular (eccentricity e${\sim}0.089$), with a period of 
${\sim}8.9\,$d \citep{Kreykenbohm+08}.
The NS orbits at a close distance from the companion's surface, 
$0.7\,$R$_{\rm B}$ at the apastron, 
and it is continuously embedded in its strong stellar wind, characterized by 
$\dot{M}_{\rm wind} = 1.5-2\times10^{-6}\,$M$_{\odot}\,$yr$^{-1}$ \citep{Watanabe+06}
and terminal velocity $v_{\rm \infty}{\sim}1100\,$km\,s$^{-1}$ \citep{Prinja+90}.
Passing within the NS accretion radius, $r_{\rm acc} = 2{GM_{\rm NS}}/{{v^2_{rel}}}$, 
this wind is accreted onto the compact object, powering the X-ray emission, with luminosity

\begin{equation}\label{eq:1}
L_{\rm X} = \frac{(GM_{\rm NS})^3\,\dot{M}_{\rm wind}}{R_{\rm NS}\,{v}_{\rm rel}^4\,D^2}\\
,\end{equation}
where $G$ is the gravitational constant, $R_{\rm NS}\sim10\,$km is the NS radius, 
$v_{\rm rel}$ the relative velocity of the wind seen from the accretion center, 
and $D$ the distance of the compact object from the center of the companion.

The observed average X-ray luminosity of ${\sim}4\times10^{36}\,$erg/s is consistent 
with a wind-fed accreting pulsar scenario (see Eq.~\ref{eq:1}).
However, it is typical of \src~to show "off-states" in which it becomes faint 
for several $10^2 - 10^3\,$s, and "giant flares", when the source flux 
suddenly increases up to a factor of $20$ \citep{Kreykenbohm+08, Sidoli+15}.
Density inhomogeneities are observed in the matter during the accretion stage 
(see \citealt{Martinez+14}, and references therein) and are considered a possible 
reason for the observed X-ray flux variability.

\begin{figure*}[t!]
\includegraphics[width=\hsize]{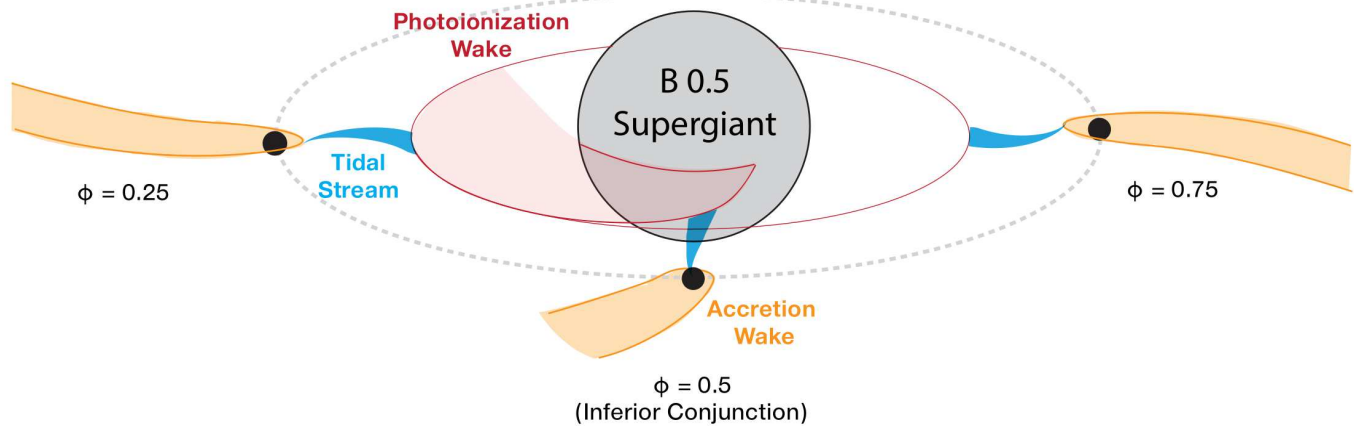}
\caption{Sketch of the binary system \src~where both radiative and gravitational effects 
	are illustrated at three different orbital phases.
	The photoionization wake is represented as a disk-like structure to show both the cumulative effect 
	along one entire orbit (the red ring) and the more localized effect (the fainter red tail) 
	trailing the NS at the inferior conjunction.
	A clumpy stellar wind, not shown in the picture, may also be taken 
	into account for a more complete scenario.
	}
\label{fig:system}
\end{figure*}

Orbital phase-resolved studies of \src, both observational and theoretical 
(see \citealt{Blondin+91}, and references therein), provide evidence of 
a strong systematic modulation of the absorbing column density with orbital phase.
This is also observed in other HMXBs (see the seminal case of Cen X--3, \citealt{Jackson75},
and more recently, the study of IGR~J$17252$--$3616$, \citealt{Manousakis+12}).
Such an effect is generally attributed to the NS influence on the stellar wind by various mechanisms.
These include the formation of a tidal stream trailing the NS (due to the almost 
filled Roche lobe of the companion), the X-ray photoionization of the wind 
(which leads to the formation of a Str{\"o}mgren sphere and a photoionization wake), 
and an accretion wake that surrounds the compact object 
(due to the focusing of the stellar wind medium by the NS gravitational influence).
A sketch of the binary system where all these structures are illustrated is shown in Fig.~\ref{fig:system}.

Many authors have reported on the X-ray light curve and spectral modulation of \src~around 
the inferior conjunction (e.g., \citealt{Eadie+75}, \citealt{Watson+77}, \citealt{Nagase89}).
At the same time, the emission from the optical companion also shows modulation with the orbital phase.
An absorption feature was reported by \citet{Smith01}, who observed a dip in the \src~UV light curve 
around the transit phase ($0.46<\phi<0.70$ in their work, where $\phi=0$ is the mid-eclipse).
Moreover, \citet{Goldstein+04}, analyzing spectroscopic \textit{Chandra} 
data taken at three different phases (i.e., eclipse $\phi=0$, $\phi=0.25$ and $\phi=0.5$), 
found an absorbed ``eclipse-like'' spectrum at $\phi=0.5$ in the soft X-ray energy range.
Similarly, \citet{Watanabe+06} found an extended structure of dense 
material that partially covers the X-ray pulsar, which is between the X-ray emitter 
and the observer at the inferior conjunction.
More recently, \citet{Naik+09} analyzed orbital phase-resolved spectra of \src~observed with RXTE, 
finding large variation of the absorbing column density along the orbital phase (up to $10^{24}\,$cm$^{-2}$), 
but their observations do not cover the whole orbit, that is, their data do not 
include observations between the inferior conjunction and the apastron.
Finally, \citet{Doroshenko+13} have shown that the observed increasing absorption at late orbital phase does not agree 
with expectations for a spherically symmetric smooth wind.

In this paper we present the spectroscopical and light curve analysis of 
\src~observed with the Monitor of All-sky X-ray Image (\maxi\,).
Our data show an excellent coverage of the entire orbit and extend over more than four years.
We extract a sample of double-peaked orbital profiles in two energy bands, whose dip is around the inferior conjunction.
We perform orbital phase-averaged and phase-resolved spectroscopy with different models 
and analyze our results using the most recent stellar wind accretion scenarios.

\label{sec:intro}

\section{Observations and data}
\begin{table}
\caption{Observation time and ephemeris used for light curves folding.}
\label{tab:setting}
\centering
\renewcommand{\arraystretch}{1.3}\begin{tabular}{c | c }
\hline\hline
Parameter & Value \\
\hline

$\Delta$T$_{Obs}$ [MJD]\tablefootmark{a} & $55133 \div 56674$ \\

P [d] & $8.964357\pm0.000029$ \\

T$_{Mid-Eclipse}$ [MJD]\tablefootmark{b} & $55134.63704$ \\

\hline

\end{tabular}\\[0.5ex]
\tablefoot{
\tablefoottext{a}{Time span of the \maxi~observations used in this work;}
\tablefoottext{b}{Taken as the phase-zero.}
}
\end{table}

The all-sky monitor \maxi~\citep{Matsuoka+09} is operating from the 
International Space Station (ISS) since August 2009.
It consists of two slit cameras working in two complementary 
energy bands: the Solid-state Slit Camera (SSC, \citealt{Tomida+11}) 
in the $0.7-10\,$keV, and the Gas Slit Camera (GSC, \citealt{Mihara+11}) 
in the $2-20\,$keV energy bands.
Here we only used archival GSC data.
The GSC module is composed of $12$ proportional counters, each of which 
consists of a one-dimensional slit-slat collimator and a proportional counter filled with Xe-gas.
Assembled, they allow covering a field of view of $1.5\times160\,$deg$^2$, observing about $75\%$ 
of the whole sky at each ISS revolution (${\sim}92\,$min), and $95\%$ per day.
The GSC typically scans a point source on the sky for $40-150\,$s during each ISS revolution 
(the actual time depends on the source incident angle for each GSC counter).
The daily flux source sensitivity at $5\sigma$ c.l. in the full energy band is about $20\,$mCrab, 
and the $1\sigma$ c.l. energy resolution is $18\%$ at $5.9\,$keV \citep{Sugizaki+11}.
These properties allowed frequently observing \src~along its 
orbit and to collecting data from this source for more than four years.

We extracted spectra and light curves of \src~from GSC data covering almost 
the whole \maxi~operation time, that is, from $55133$ to $56674\,$MJD (see Table~\ref{tab:setting}).
Our data were properly corrected to account for the time-dependent effective area of the cameras.
All extracted spectra were rebinned to contain at least 200 photons per energy bin.
Following the \maxi~team recommendation, we added a systematic error at a level of $1\%$ 
to the final count rates of the extracted spectra.

\label{sec:observation}

\section{Orbital profiles samples analysis}
\begin{figure}[t!]
\includegraphics[width=\hsize]{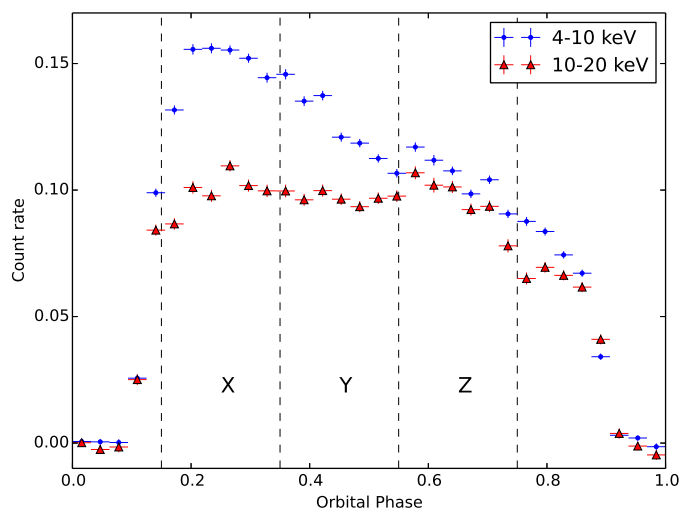}
\caption{\src~folded orbital light curves in $4-10$ (blue dots), and $10-20\,$keV (red triangles) energy ranges.
	The horizontal bars indicate the phase-bin.
	The mid-eclipse time is taken as the phase zero.
	The three orbital phase bins (X, Y, and Z) used for the phase-resolved spectroscopy 
	of the double-peaked sample are indicated (see text).
	(A color version of this figure is available in the online journal).}
\label{fig:profiles}
\end{figure}

The mission-long light curve of \src~as observed by \maxi/GSC 
is provided by the \maxi~team\footnote{The entire \maxi/GSC \src~light curve 
can be explored at \url{http://maxi.riken.jp/top/index.php?cid=1&jname=J0902-405}}
at different energy ranges.
It exhibits a rich variety of timing phenomena.
However, we here focus on the orbital properties of the source.
We therefore extracted \maxi/GSC light curves with a time resolution of 
$6\,$h in two energy bands, $4-10$ and $10-20\,$keV.
Then, we folded the light curves to produce average orbital profiles in 
the two chosen energy bands, using the orbital ephemeris from \citet{Kreykenbohm+08} 
that we report in Table~\ref{tab:setting}.
The phase zero is taken at the mid-eclipse time (see Fig.~\ref{fig:sketch}).
The resulting orbital light curves are shown in Fig.~\ref{fig:profiles}, 
where the strongest changes are observed in the softer band light curve.
This is explained as strong absorption column modulation along the binary orbit.

Successively, we considered the difference between individual orbital 
light curves (defined as the members of the sample) and the average orbital 
profile (defined as the template).
To obtain a smooth template of the two folded light curves, we interpolated with 
a cubic spline function between all the pairs of consecutive data points 
(the knots) of the light curves.

\begin{figure}[t!]
\includegraphics[width=\hsize]{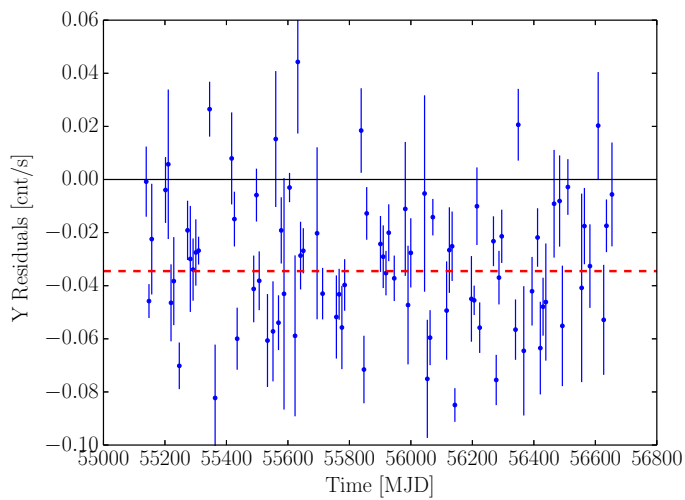}
\caption{
	Residuals of single orbital profiles from the template in the $4-10\,$keV 
	energy band for the Y phase-bin (see text).
	The residuals are compared to the observation time on the x-axis.
	The thin black line is the zero-value, while the red dashed line is the $30\%$ 
	limit below which the selection criterion in the Y phase-bin is fullfilled.
	Error bars correspond to the sample standard deviation 
	($\sigma/\sqrt{N}$, with $N$ number of data points).
	Not all the profiles below the red dashed line are selected as double-peaked 
	profiles because of the conditions imposed on the X and Z phase-bins.
	}
\label{fig:3D}
\end{figure}

\begin{figure}[!t]
\includegraphics[width=\hsize]{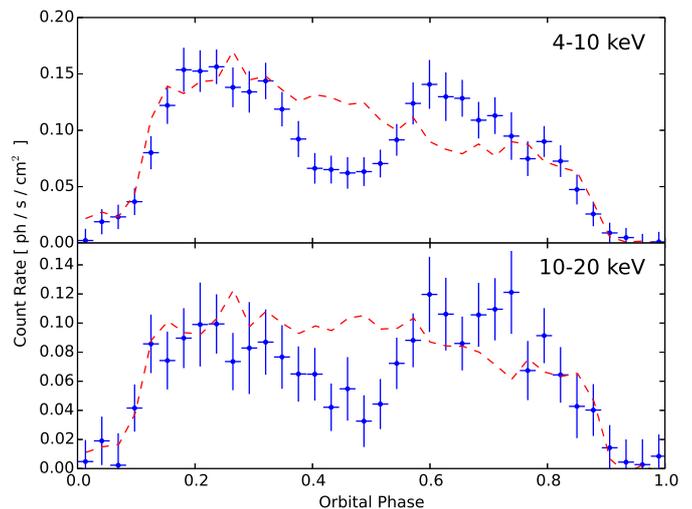}
\caption{Average orbital profiles of the double-peaked and of the standard samples 
	in $4-10\,$keV (top) and $10-20\,$keV (bottom).
	The blue dots are the data points obtained by stacking the memebers with large 
	residuals in the Y phase bin (see text).
	The vertical error bars correspond to the standard deviation of the average of the 
	count rates in each phase bin, while the horizontal bars indicate the phase-bin width.
	The red dashed lines represents the function interpolated between the data points of 
	the standard samples in each energy band.
	}
\label{fig:double_profiles}
\end{figure}

From our line of sight, the accretion wake effects are more prominent for about $0.2$ 
in orbital phase (i.e., about $1.5\,$days) around the inferior conjunction (see Fig.~\ref{fig:sketch}).
To properly sample these effects, we defined three orbital phase bins that we call X, Y, and Z, 
corresponding to phase intervals $0.15-0.35$, $0.35-0.55$, and $0.55-0.75$, 
respectively (see Fig.~\ref{fig:profiles}).
For analogous reasons we imposed for each member a minimum number of data points 
in each phase bin equal to $4$ (i.e., $1\,$d observation), 
which leaves us with a total of $84$ statistically acceptable cycles.
This allowed us to perform a systematic analysis of those members that differ from 
the template in each of the three defined phase bins.
For the systematic comparison of the members’ profile against the template, 
we considered only the relative shape, avoiding any bias from flux variation.
For this reason, we normalized
each member with a normalization factor equal to the ratio of the average count rate 
between $0.10-0.85$ and the template average count rate in the same phase interval.
Successively, we subtracted the count rates of the template from the count rates of 
each member in the X, Y, and Z phase bins.
To evaluate the template and the member count rate in each phase bin, we calculated 
the median of the data points.
The obtained count rate difference is referred to as the residuals 
with respect to the template.
The residuals of the members from the template for the Y phase-bin in the 
$4-10\,$keV energy band are shown in Fig.~\ref{fig:3D}.
This plot shows a scattered distribution of the residuals, as expected because of 
the high variability of the source.
However, the plot also clearly shows that residuals tend to cluster toward negative values.
Therefore, we performed a new analysis and stacked only those members 
that had high ($>30\%$) negative residuals (i.e., $<-0.0345\,$cnt/s) in the Y phase bins 
and at the same time low ($<30\%$) residuals in the X and Z phase bins.
This allowed us to obtain a new average orbital profile in the $4-10\,$keV 
energy band, made of $12$ out of $84$ total orbital cycles (i.e., about $15\%$ 
of the analyzed cycles).
This is shown in Fig.~\ref{fig:double_profiles}.
Using a different definition of the residuals, where the ratio between 
the median values is taken (instead of their difference), shows the very same trend 
as that in Fig.~\ref{fig:3D}, and leads to the same selected profiles.

Since our normalization procedure could remove flux-dependent features, 
we carefully verified that these selected profiles were not associated with a specific 
luminosity level before normalization.
In addition, for all of these $12$ selected profiles, the null hypothesis was tested 
against the template with normality tests ($\chi^2$, Kolmogorov-Smirnov, Shapiro-Wilk) 
and was rejected at a significance level higher than $3\sigma$.
Furthermore, they all show a median value in the Y phase-bin that deviates 
at least $3\sigma$ from the template.
These profiles are characterized by a double-peak shape and a flux drop 
around the inferior conjunction.
We refer to them as the double-peaked sample.

Similar results are obtained if the same procedure is performed in the $10-20\,$keV energy band.
In this case, the selected profiles are $11$ out of $84$.
The overall profile shape is very similar to the profile obtained in the $4-10\,$keV energy band,
with a double-peak morphology and a dip around the inferior conjunction.
The extracted sample in the $10-20\,$keV energy band is shown in Fig.~\ref{fig:double_profiles}.

Finally, we extracted another independent sample in the two energy bands by 
stacking all the statistically acceptable orbital profiles except for the 
double-peaked members in each respective energy band.
We refer to this sample as the standard sample.
The standard sample orbital profiles in the two energy bands are shown with dashed lines in Fig.~\ref{fig:double_profiles}.

\label{sec:Population}

\section{Phase-averaged spectral analysis}

The orbital phase-averaged spectral analysis of \src~was performed separately 
for the two samples defined in Sect.~\ref{sec:Population}.
Because the physical problem is complex, a fully physical model of the X-ray production 
mechanism in accreting X-ray pulsars has not been implemented so far.
Therefore, an empirical model has to be used.
Following the common empirical model used in the literature (see, e.g., \citealt{Schanne+07}), 
we adopted a cutoff power-law model (\textit{cutoffpl} in XSPEC)
to which a blackbody component was added to account for soft excess 
(see \citealt{Hickox+04} and references therein).
On the other hand, to account for the expected inhomogeneous structure of the ambient wind in \src,
we also tested the partial covering model (\textit{pcfabs*powerlaw} in XSPEC)
characterized by a local absorbing component ($N_{\rm H}^{pc}$) that affects 
only a fraction $f$ of the original spectrum.
This model has also been successfully used to fit \textit{NuStar} data of \src~\citep{Fuerst+14}.
To both models, photoelectric absorption by neutral matter \citep{Morrison+McCammon83}
was added, to account for the absorption column density ($N_{\rm H}$) 
along the entire line of sight.
We assumed solar abundances by \citet{Anders+Grevesse89} for this.
Using the metal abundances of \citet{Kaper+93} for HD~$77581$ 
(i.e., sub-solar C, and super-solar N, Al)
does not affect our results, which are well consistent within $1\sigma$.
Therefore, we used solar abundances throughout the rest of the analysis.
We also added two iron Gaussian line components \citep{Odaka+13}, K$\alpha$ and 
K$\beta$ at $6.4$ and $7.08\,$keV..
The same components were included in the models used for the phase-resolved spectroscopy.
The results of the fits are given in Table~\ref{tab:averaged}.
Both models return acceptable fits for the standard and the double-peaked sample.
However, for the standard sample, the partial covering model describes the data not as well as the cutoff power-law model.


\label{sec:spectral}

\section{Phase-resolved spectral analysis}

\subsection{Orbital phase-resolved spectroscopy of the double-peaked sample}\label{subsec:double_prs}

To investigate the nature of the dip in the double-peaked sample,
we extracted X-ray spectra of \src~in the X, Y and Z orbital phase bins.
For this, we filtered \maxi/GSC data with GTIs corresponding to the 
three phase bins in the double-peaked profiles (see Sect.~\ref{sec:Population}).
Based on the results from Sect.~\ref{sec:spectral}, we used the same models as were 
applied in the phase-averaged case to fit the orbital phase-resolved spectra:
(a) a cutoff power-law model plus a blackbody component, and (b) a partial covering model.
To avoid unphysical best-fit solutions when model (a) was used,
we fixed the cutoff energy and the blackbody temperature
to their respective phase-averaged values.
For model (a) we also compared the results for a fixed (to the phase-averaged value, 
$\Gamma=0.27$, see Table~\ref{tab:averaged}) or free photon index.
We found that the model with free photon index is statistically preferable ($\Delta\chi^2>20$).
Furthermore, fitting the Y phase bin spectrum, the model with a fixed photon index leads 
to unphysically high values of the blackbody normalization (to compensate for the soft excess).
Thus, we did not consider the cutoff power-law model with a fixed photon index a viable 
model in our analysis (similar considerations hold for the phase-resolved spectroscopy of 
the standard sample, see Sect.~\ref{subsec:standard_prs}).
It is often possible to find a statistically equivalent solution for model (b) 
where the partial covering component is not necessary.
We tested the significance of the partial covering component with the F-test,
including such a component only when it was significant at a $99\%$ c.l.
Otherwise, we preferred the solutions where the partial covering component was not included.
The same criterion was adopted for the standard sample case (see Sect.~\ref{subsec:standard_prs})

\begin{figure}[t!]
\includegraphics[width=\hsize]{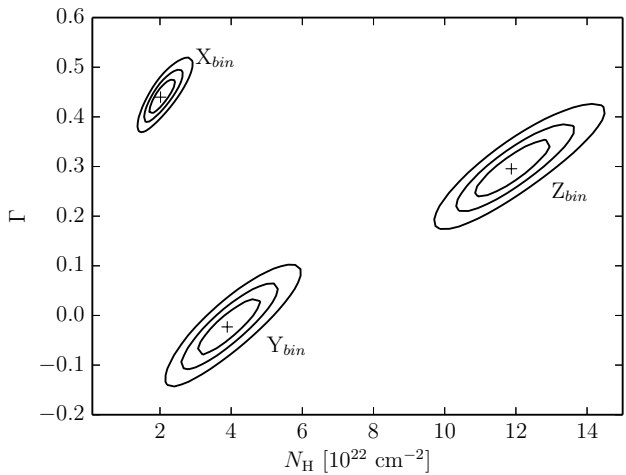}
\caption{$\chi^2$ contour plots for the \src~double-peaked sample fitted 
	with an absorbed cutoff power-law model including a blackbody component (model \textit{a}, see Table~\ref{tab:doubleprs}).
	The ellipses are $\chi^2$-contours for two parameters ($N_{\rm H}$ and $\Gamma$).
	The contours correspond to $\chi^{2}_{\rm min} + 1.0$ (the projections of this contour to the parameter 
	axes correspond to the $68\%$ uncertainty for one parameter of interest), $\chi^{2}_{\rm min} + 2.3$ 
	($68\%$ uncertainty for two parameters of interest), and $\chi^{2}_{\rm min} + 4.61$ ($90\%$ uncertainty 
	for two parameters of interest). The respective orbital phase bins are indicated.}
\label{fig:contoursdouble}
\end{figure}

Models (a) and (b) both return a good fit, and the results are shown in Table~\ref{tab:doubleprs}.
Model (a) shows orbital modulation of the spectral photon index $\Gamma$.
To better explore the spectral variation, we produced $\chi^2$-contour plots of 
$\Gamma$ and $N_{\rm H}$ in the three phase bins (see Fig.~\ref{fig:contoursdouble}).

\begin{figure}[!t]
\includegraphics[angle=270, width=\hsize]{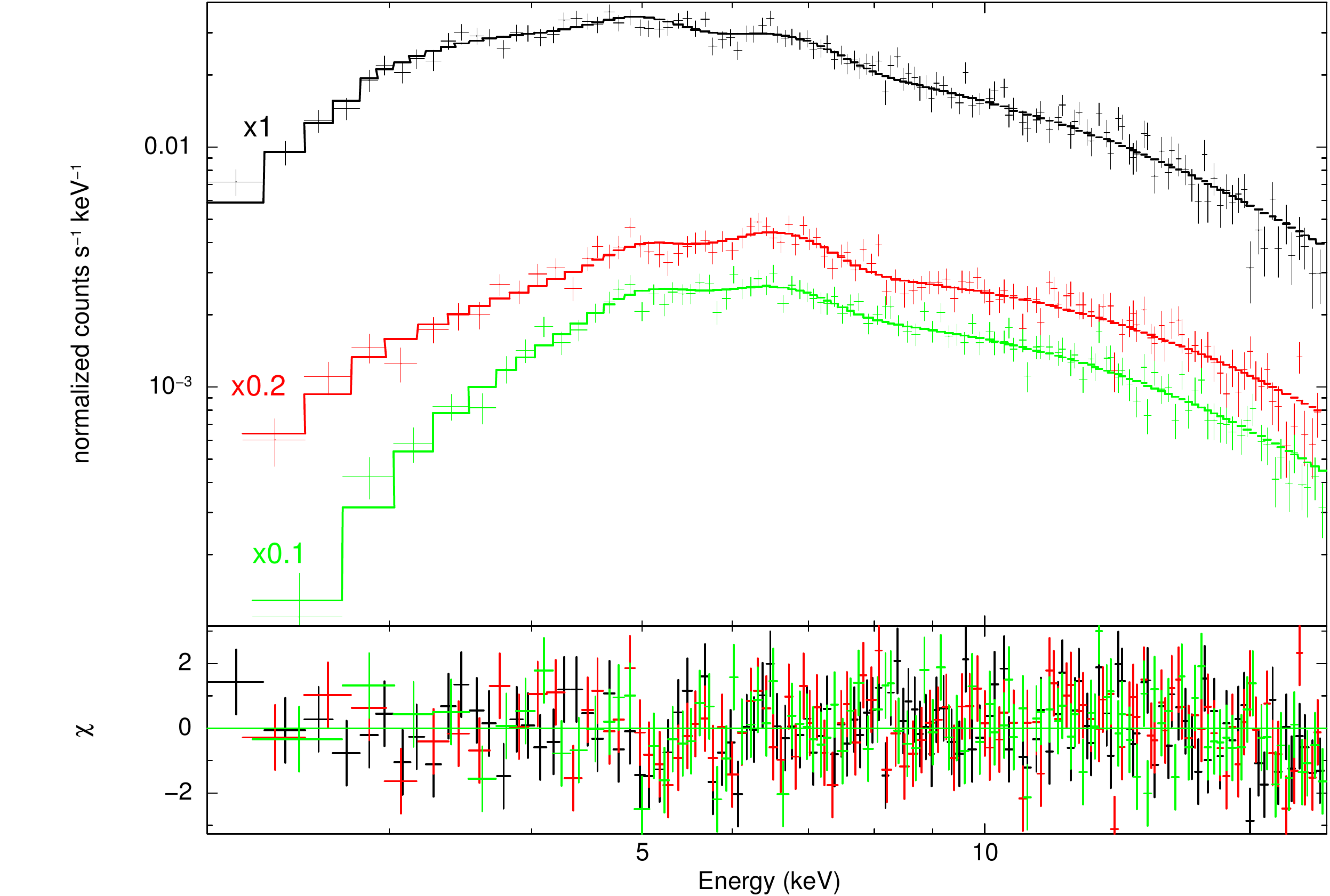} 
\caption{Orbital phase-resolved spectra of the double-peaked sample fitted with 
	a partial covering model, when significant (see text).
	The top panel shows from top to bottom data and folded model of the three spectra 
	in phase bins X, Y,  and Z.
	The spectra have been multiplied by the factor shown above each curve 
	on the left side for better visualization.
	The bottom panel shows the residuals for the best-fit model.
	See Table~\ref{tab:doubleprs} for all spectral parameters.
	}
\label{fig:double_sim_shifted}
\end{figure}

Conversely, model (b) does not show photon index modulation along the orbital phase,
although it shows variation of the column density.
A plot of the double-peaked phase-resolved spectra fit with the partial covering 
model is shown in Fig.~\ref{fig:double_sim_shifted}.

\subsection{Orbital phase-resolved spectroscopy of the standard sample}\label{subsec:standard_prs}

\begin{table*}
\caption{Best-fit spectral parameters of the orbital phase-averaged spectra from the double-peaked and standard samples. The errors correspond to the $1\sigma$-uncertainties.}
\label{tab:averaged}
\centering
\renewcommand{\arraystretch}{1.3}\begin{tabular}{l | c c | c c }
\hline\hline
\multicolumn{0}{l|}{Model}        & \multicolumn{2}{|c|}{\textbf{Cutoff PL + BB}}   & \multicolumn{2}{|c|}{\textbf{Part. Cov.}} \\[0.5ex]
\hline
Sample                            & Double-peaked          & Standard               & Double-peaked          & Standard          \\
\hline

$N_{\rm H}/10^{22}\,$cm$^{-2}$          & $3.91_{-0.36}^{+0.43}$ & $4.29_{-0.30}^{+0.31}$ & $2.55_{-2.55}^{+2.18}$ & $2.02_{-0.79}^{+0.68}$ \\

$N_{\rm H}$\tablefootmark{pc}$/10^{22}\,$cm$^{-2}$&  --          &  --                    & $14.9_{-5.4}^{+16.0}$ & $12.7_{-1.7}^{+2.5}$ \\

Cov. Fract.\tablefootmark{pc}     &  --                    &  --                    & $0.63_{-0.18}^{+0.21}$ & $0.64_{-0.06}^{+0.07}$ \\

$\Gamma$                          & $0.27_{-0.03}^{+0.04}$ & $0.48_{-0.06}^{+0.06}$ & $1.09_{-0.07}^{+0.10}$ & $1.13_{-0.03}^{+0.03}$ \\

$E_{\rm cut}\,$[keV]                  & $15.3_{-3.4}^{+0.4}$   & $20.1_{-2.0}^{+2.5}$   &  --                    &  --                    \\

$kT_{\rm BB}\,$[keV]                  & $0.16_{-0.04}^{+0.04}$   & $0.14_{-0.08}^{+0.06}$ &  --                  &  --                    \\



$\chi^2/${\scriptsize{D.o.F.}}$ = \chi^2_{\rm red}$ 
                                  & $285/311 = 0.92$       & $393/351 = 1.12$       & $318/311 = 1.02$       & $480/352 = 1.36$     \\
\hline
\end{tabular}
\tablefoot{
\tablefoottext{pc}{Partial Covering Model.}
}
\end{table*}

\begin{table*}
\caption{Best-fit spectral parameters of the phase-resolved spectra from the double-peaked sample. The errors correspond to the $1\sigma$-uncertainties.}
\label{tab:doubleprs}
\centering
\renewcommand{\arraystretch}{1.3}\begin{tabular}{l | c c | c c | c c }
\hline\hline
\multicolumn{0}{l|}{Phase-bins} & \multicolumn{2}{|c|}{\textbf{X}}                  & \multicolumn{2}{|c|}{\textbf{Y}}                 & \multicolumn{2}{|c}{\textbf{Z}}\\[0.5ex]
\hline
Model                             & Cutoff PL + BB         & Part. Cov.             & Cutoff PL + BB          & Part. Cov.             & Cutoff PL + BB         & Part. Cov. \\
\hline

$N_{\rm H}/10^{22}\,$cm$^{-2}$            & $2.02_{-0.31}^{+0.41}$ & $3.68_{-2.16}^{+0.29}$ & $3.89_{-0.87}^{+0.93}$  & $2.49_{-2.49}^{+2.07}$ & $11.9_{-1.0}^{+1.1}$   & $16_{-3}^{+1}$ \\

$N_{\rm H}$\tablefootmark{pc}$/10^{22}\,$cm$^{-2}$ & --     & -- & --                      & $16.1_{-5.3}^{+11.3}$  & --                     & -- \\

Cov. Fract.\tablefootmark{pc}      & --                 & --         & --    & $0.64_{-0.09}^{+0.17}$ & --                     & -- \\

$E_{\rm cut}\,$[keV]                  & $15.3$ (fixed)         & --                     & $15.3$ (fixed)          & --                     & $15.3$ (fixed)         & -- \\

$\Gamma$                          & $0.44_{-0.03}^{+0.04}$ & $1.12_{-0.04}^{+0.02}$ & $-0.02_{-0.06}^{+0.06}$ & $0.91_{-0.11}^{+0.29}$ & $0.29_{-0.06}^{+0.06}$ & $1.10_{-0.05}^{+0.05}$ \\

Norm$_{\rm PL}$            & $0.110_{-0.008}^{+0.009}$ & $0.235_{-0.017}^{+0.014}$ & $0.003_{-0.005}^{+0.005}$ & $0.129_{-0.038}^{+0.151}$ & $0.089_{-0.01}^{+0.015}$ & $0.256_{-0.031}^{+0.042}$ \\

kT$_{\rm BB}\,$[keV]\tablefootmark{b} & $0.16$ (fixed)         & --                     & $0.16$ (fixed)          & --                     & $0.16$ (fixed)         & -- \\

 
$\chi^2/${\scriptsize{D.o.F.}}$ = \chi^2_{\rm red}$ 
                                  & $113/121{=}0.93$       & $121/120{=}1.00$       & $113/108{=}1.05$        & $117/107{=}1.09$       & $101/102{=}0.99$       & $113/101{=}1.11$ \\

\hline\hline
\end{tabular}
\tablefoot{
\tablefoottext{pc}{Partial Covering Model.}
\tablefoottext{b}{Blackbody component.}
}
\end{table*}

\begin{figure}[!t]
\includegraphics[width=\hsize]{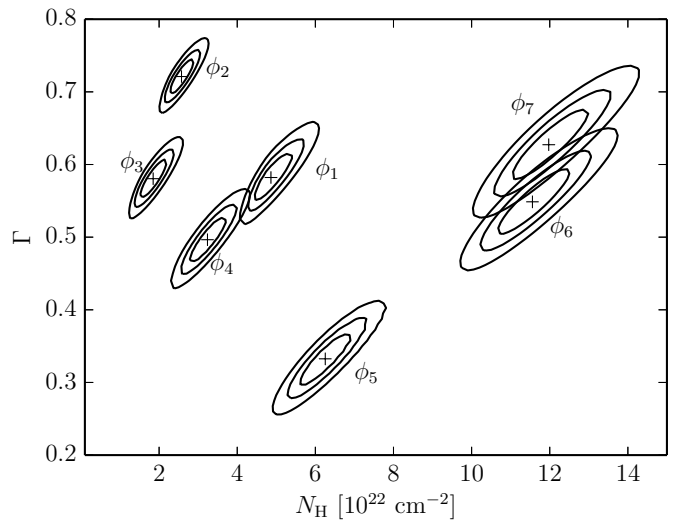}
\caption{$\chi^2$ contour plots for the \src~standard sample fitted with an absorbed cutoff power-law model 
	including a blackbody component (see Table~\ref{tab:prs_standard}).
	Symbols are the same as in Fig.~\ref{fig:contoursdouble}.
	The respective orbital phase bins are indicated.}
\label{fig:contoursaverage}
\end{figure}

Similarly to the double-peaked sample, we performed orbital phase-resolved 
spectroscopy of the standard sample.
However, for this sample, the statistics is much higher than in the previous case, 
so that a finer binning is possible.
We divided the standard sample profile into seven phase bins, each one wide $0.1$ in phase, from $0.1$ to $0.8$.
Phase bins around the eclipse time are not suitable for our purpouses 
and suffer from lower statistics, therefore they were not analyzed in our work.
The analyzed phase bins are labeled $\phi_{\rm i}$, with $i=1, ..., 7$.
Then we filtered our data with GTIs corresponding to the seven orbital phase bins 
and extracted orbital phase-resolved spectra of the standard sample.
Based on the results from Sect.~\ref{sec:spectral}, we fit the phase-resolved spectra 
with the same two models: (a) a cutoff power-law with a blackbody component, 
and (b) a partial covering model.
To avoid unphysical best-fit solutions, we fixed the cutoff energy and the blackbody temperature 
of model (a) to their respective phase-averaged value.
As already mentioned in Sect.~\ref{subsec:double_prs}, model (b) returns two statistically 
equivalent solutions: one where the partial covering component is statistically significant,
and the other where the same component is not necessary.
In such cases, we adopted the solution in which the partial covering component 
was not required, unless it was significant at a $99\%$ c.l.
The results are given in Table~\ref{tab:prs_standard}.
All phase-resolved spectra return good fits for each of the two tested models.
However, the cutoff power-law model with blackbody shows a marked modulation of the 
spectral photon index $\Gamma$ along the orbital phase.
The trend of this modulation is similar to that observed for the double-peaked sample 
(see Sect.~\ref{subsec:double_prs}).
For a deeper investigation of these results, $\chi^2$-contour plots of $\Gamma$ and $N_{\rm H}$ 
were produced (see Fig.~\ref{fig:contoursaverage}).

\begin{figure}[t!]
\includegraphics[angle=270, width=\hsize]{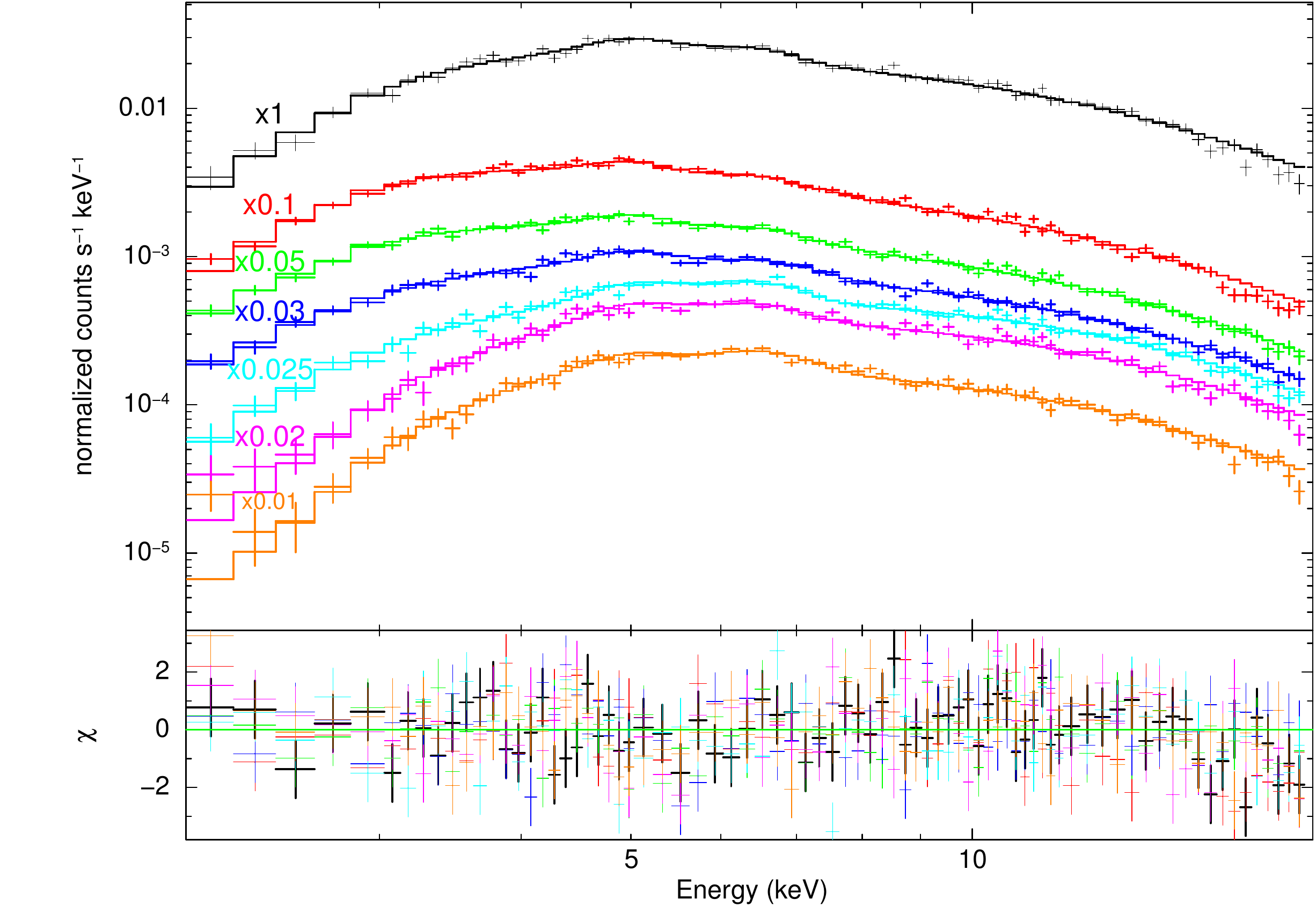}
\caption{
	Orbital phase-resolved spectra of the standard sample fit 
	with a partial covering model, when significant (see text).
	The top panel shows from top to bottom data and folded model 
	of the spectra in phase bins from $\phi_1$ to $\phi_7$.
	The bottom panel shows the residuals for the best-fit model.
	Symbols are the same as in Fig.~\ref{fig:double_sim_shifted}.
	See Table~\ref{tab:prs_standard} for all spectral parameters.
	}
\label{fig:standard_sim_shifted}
\end{figure}

The partial covering model shows only marginal variation of the photon index $\Gamma$,
while the column density clearly changes along the orbital phase,
and the covering fraction is consistent with a constant value.
A plot of the standard sample phase-resolved spectra fit with the partial covering 
model is shown in Fig.~\ref{fig:standard_sim_shifted}.
\begin{sidewaystable*}
\caption{Best-fit spectral parameters of the orbital phase-resolved spectra from the standard sample. The errors correspond to the $1\sigma$-uncertainties.}
\label{tab:prs_standard}
\centering
\renewcommand{\arraystretch}{1.3}\begin{tabular}{l | c | c | c | c | c | c | c}
\hline\hline
\multicolumn{0}{l|}{Sample} & \multicolumn{7}{|c|}{\textbf{Cutoff PL + BB} ($E_{cut} = 20.1\,$keV fixed, $kT_{BB}\tablefootmark{b} = 0.14\,$keV fixed)} \\[0.5ex]
\hline
Phase-bin                   & $\phi_1$                 & $\phi_2$                  & $\phi_3$                  & $\phi_4$                  & $\phi_5$                  & $\phi_6$ & $\phi_7$ \\[0.5ex]
\hline

$N_{\rm H}/10^{22}$\,cm$^{-2}$    & $4.86_{-0.42}^{+0.55}$   & $2.57_{-0.29}^{+0.31}$   & $1.85_{-0.33}^{+0.35}$    & $3.24_{-0.46}^{+0.48}$    & $6.25_{-0.64}^{+0.68}$    & $11.5_{-0.8}^{+0.9}$            & $11.9_{-0.9}^{+1.0}$\\ 

$\Gamma$                    & $0.58_{-0.03}^{+0.03}$   & $0.72_{-0.02}^{+0.02}$    & $0.58_{-0.03}^{+0.03}$    & $0.49_{-0.03}^{+0.03}$   & $0.33_{-0.03}^{+0.04}$    & $0.55_{-0.04}^{+0.04}$          & $0.63_{-0.05}^{+0.05}$\\

Norm$_{\rm PL}$ & $0.111_{-0.008}^{+0.009}$ & $0.194_{-0.009}^{+0.010}$& $0.126_{-0.007}^{+0.007}$& $0.095_{-0.006}^{+0.007}$& $0.066_{-0.005}^{+0.005}$& $0.11_{-0.01}^{+0.01}$ & $0.12_{-0.01}^{+0.01}$ \\

$\chi^2/${\scriptsize{D.o.F.}}$ = \chi^2_{\rm red}$ 
                            & $173/212=0.82$         & $233/247 = 0.94$          & $244/237 = 1.03$          & $216/228 = 0.95$          & $246/228 = 1.08$          & $216/205 = 1.05$ & $204/199 = 1.03$ \\

\hline\hline
\multicolumn{0}{l|}{Model} & \multicolumn{7}{|c|}{\textbf{Partial Covering}} \\[0.5ex]
\hline
Phase-bin                         & $\phi_1$               & $\phi_2$               & $\phi_3$               & $\phi_4$               & $\phi_5$               & $\phi_6$               & $\phi_7$ \\[0.5ex]
\hline

$N_{\rm H}/10^{22}$\,cm$^{-2}$            & $6.3_{-1.0}^{+0.3}$ & $1.7_{-1.7}^{+1.2}$ & $0.^{+1.4}$ & $2.2_{-1.5}^{+0.93}$ & $5.6_{-0.6}^{+0.9}$ & $13.3_{-1.7}^{+0.7}$   & $13.7_{-1.7}^{+0.7}$\\ 

$N_{\rm H}$\tablefootmark{pc}$/10^{22}\,$cm$^{-2}$ & -- & $7.1_{-2.0}^{+7.9}$ & $6.4_{-1.0}^{+4.4}$ & $13_{-4}^{+8}$   & $32_{-8}^{+10}$ & -- & --\\ 

Cov. Fract.\tablefootmark{pc}      & -- & $0.49_{-0.24}^{+0.29}$ & $0.68_{-0.25}^{+0.02}$ & $0.51_{-0.10}^{+0.19}$ & $0.61_{-0.03}^{+0.05}$ & -- & -- \\

$\Gamma$                          & $1.09_{-0.02}^{+0.03}$ & $1.26_{-0.03}^{+0.05}$ & $1.13_{-0.04}^{+0.05}$ & $1.12_{-0.06}^{+0.08}$ & $1.23_{-0.19}^{+0.18}$ & $1.08_{-0.04}^{+0.04}$ & $1.15_{-0.04}^{+0.04}$\\

Norm$_{\rm PL}$ & $0.21_{-0.01}^{+0.01}$ & $0.40_{-0.04}^{+0.06}$& $0.26_{-0.02}^{+0.03}$& $0.26_{-0.04}^{+0.07}$& $0.39_{-0.17}^{+0.27}$& $0.22_{-0.02}^{+0.02}$ & $0.23_{-0.02}^{+0.02}$ \\

$\chi^2/${\scriptsize{D.o.F.}}$ = \chi^2_{\rm red}$ 
                                  & $192/211 = 0.91$       & $250/247 = 1.01$       & $248/236 = 1.05$       & $212/227 = 0.94$       & $251/227 = 1.11$       & $237/204 = 1.16$       & $219/198 = 1.11$ \\

\hline
\end{tabular}
\tablefoot{
\tablefoottext{b}{Blackbody component.}
\tablefoottext{pc}{Partial Covering Model.}
} \\[6.ex]
\end{sidewaystable*}

\label{sec:analysis:prs}

\section{Discussion}
The main observational results of our work can be summarized as follows:
\begin{itemize}
\item About $15\%$ of the orbital light curves in \src~shows a double-peaked orbital profile, 
      with a dip around the inferior conjunction, both in the $4-10\,$keV and $10-20\,$keV energy bands;

\item orbital phase-resolved spectra of both the double-peaked and the standard sample are well fit by 
      a cutoff power-law model and by a partial covering model;

\item the cutoff power-law model leads to orbital modulation of both 
	the spectral photon index and the column density;

\item a partial covering model fits the data equally well.
      No orbital modulation of the spectral photon index is observed in this case.
      In contrast, orbital modulation of the column density is observed.

\end{itemize}

In the following we discuss our results within the context of the most recent 
wind accretion scenarios that have been proposed for \src.

\subsection{Double-peaked orbital light curve}\label{subsec:doublelc}

Indications of a double-peaked orbital light curve in Vela~X-1 have already been reported
by earliest observations of the source with \textit{Ariel V} \citep{Eadie+75, Watson+77}.
Based on \textit{Tenma} data, \citet{Nagase89} found an increase of the 
column density along the NS orbit, in particular after the transit phase.
\citet{Blondin+91} took both gravitational and radiative effects into account and 
reproduced the observed orbital phase dependence of the column density.
They found that an accretion wake that extends toward the observer's line 
of sight at the inferior conjunction produces a local peak in the 
column density absorption along the binary orbit.

Intriguing evidence on the accretion wake in \src~was obtained by \citet{Smith01},
who inferred that a hot source around the transit phase of the NS is present by 
analyzing He\,$\lambda1640$ absorption line strengths.
Based on \textit{Chandra} X-ray data, \citet{Watanabe+06} also have inferred the presence 
of a dense cloud obscuring the compact source at the inferior conjunction.
All these results are tightly related to the observational features reported in this work.

The sample of double-peaked orbital profiles, obtained from the very long 
\textit{MAXI/GSC} monitoring of Vela X-1, shows a remarkable dip at phase Y, 
also in the $10-20\,$keV band. 
This is challenging to explain solely by invoking absorption from neutral matter.
The X-ray flux observed at the dip phase is about half of the flux in the 
standard sample at the same phase (see Fig.~\ref{fig:profiles}).
To halve a luminosity of ${\sim}4\times10^{36}\,$erg/s in the $10-20\,$keV band, 
a column density value of $N_H\,{\sim}\,2\times10^{24}\,$cm$^{-2}$ is needed,
${\sim}10$ times higher than what is observed in our study.

In addition, hydrodynamical simulations of stellar winds in HMXBs 
(\citealt{Blondin+91}, \citealt{Manousakis+12}) predict a local peak of the 
column density $N_{\rm H}$ around the inferior conjunction on the order of 
$10^{23}\,$cm$^{-2}$.
This agrees with our observations.
We also mention that the \inte~analysis of \citet{Fuerst+10} highlights a possible 
dip-like feature around the transit in the phase-resolved flux histograms of \src~at 
$20-60\,$keV, that is, at an energy band that is almost unaffected by absorption.
Our conclusion is therefore that some additional mechanism must be at work 
that contributes to the flux reduction at the inferior conjunction, 
and that is triggered only in a limited sample of orbital cycles.

A contribution to the observed reduced flux could come from the Compton scattering 
at the low-energy limit, where the cross section is characterized by only a weak dependence 
on the photon energy and can be approximated by the Thomson scattering cross section 
($\sigma_T=6.65\times10^{-25}\,$cm$^{2}$).
To efficiently scatter X-rays from the accreting pulsar, 
the surroundings of the compact object need to be ionized and dense enough.
Hydrodynamical 2D and 3D simulations of the stellar wind in \src~by \citet{Mauche+08} showed that 
the interaction of the wind with the NS leads to strong variations of the wind parameters.
These authors found that the accretion wake reaches high temperatures (${\sim}10^8\,$K) 
and a high ionization parameter ($\xi=L_x / (n\,R_{\rm ion}^2)\sim10^4\,$erg\,cm\,s$^{-1}$,
where $n$ is the number density at a given point, while $R_{\rm ion}$ 
is the distance from this point to the X-ray source).
The accretion wake extends to scales on the order of the optical companion size 
($R\sim10^{12}\,$cm) with a number density of ${\sim}10^{9-12}\,$cm$^{-3}$,
which, taking an average value of $n\sim10^{10}\,$cm$^{-3}$,
leads to a column density of ${\sim}10^{23}\,$cm$^{-2}$.
This is similar to the values obtained from our spectral analysis 
of the double-peaked sample in the Y phase bin 
(see Sect.~\ref{subsec:double_prs_results} and Table~\ref{tab:doubleprs}). 
Moreover, the bremsstrahlung cooling time is about $10^{11}\sqrt{T}/n$~s 
\citep{Rybicki+Lightman79} that is, ${\sim}10^5\,$s, or about one day, 
which agrees with the observed time duration of the dip.
In other words, the gas stays ionized for a time long enough to allow Thomson scattering.
The ionized gas is not traced by the spectral absorption in the soft X-ray band.
Nonetheless, its electron density $n$ is high enough to result in an optical depth value of 
$\tau=n\,l\,\sigma_T=0.5-0.7$ (where $l=2\times10^{12}\,$cm is the length of the accretion wake).
Thus, when the ionized stream of the accretion wake passes through the observer's direction, 
that is, during the transit, a substantial part of the X-ray radiation can be efficiently 
scattered out of our line of sight.
This scenario is further supported by the normalization of the power-law component,
which appears to be modulated along the orbit (see Table~\ref{tab:doubleprs}),
indicating an intrinsically dimmer flux at the inferior conjunction
(although large parameter uncertainties need to be considered).

Stellar-wind variability may then be responsible for the modifications of the accretion wake,
which in turn modify the orbital light curve shape, alternating between a double-peaked and a standard morphology.
We therefore conclude that the ionized gas in the hot accretion wake,
combined with its high column density, is a possible cause for the dip that is observed 
at the inferior conjunction in the double-peaked sample light curves.

\subsection{Spectral results}

\subsubsection{Orbital phase-averaged spectroscopy}\label{subsec:phase_average_results}

Orbital phase-averaged spectra  in the $2-20\,$keV energy range 
of both the double-peaked and the standard sample were modeled 
with a cutoff power-law model, adding a blackbody component to account for the soft excess,
and with partial covering (Table~\ref{tab:averaged}).
For the double-peaked sample, both models return a good fit.
For the standard sample the partial covering model is only marginally acceptable, 
while the cutoff power-law describes the data well.

As a result of the inhomogeneous environment in which the NS travels along the binary orbit,
our phase-averaged results are not straightforward in terms of physical interpretation.
Instead, these results were used to avoid unphysical best-fit solutions 
of the orbital phase-resolved spectra, fixing some of the model parameters 
to their phase-averaged value (see Sect.~\ref{subsec:double_prs_results} 
and Sect.~\ref{subsec:standard_prs_results}).

\subsubsection{Double-peaked sample phase-resolved spectroscopy}\label{subsec:double_prs_results}

Fitting the phase-resolved spectra of the double-peaked sample
with a cutoff power-law results in orbital modulation of the photon index 
(see Table~\ref{tab:doubleprs}), which has never been observed in \src~to the best of our knowledge.
The spectrum becomes harder at the inferior conjunction (Y phase bin) 
than at earlier and later phases (X and Z phase bins, respectively).
Contour plots for $N_{\rm H}$ and $\Gamma$ (shown in Fig.~\ref{fig:contoursdouble})
appear clearly distinct from one phase bin to another,
indicating that the modulation is not due to the intrinsic coupling between the two parameters.

\citet{Prat+08} reported photon index modulation 
along the binary orbit of HMXB IGR~J$19140$+$0951$.
These authors ascribed the observed orbital modulation to a variable visibility
of the region that produces the soft excess.
Since we do not observe any modulation of the blackbody component,
our study does not support this interpretation.
To the best of our knowledge, we did not find any reliable physical interpretation for this phenomenon.

Modeling with partial covering does not show an orbital modulation of the photon index.
This model results in absorption increase along the orbital phase 
(see Sect.~\ref{subsec:standard_prs_results} for a more comprehensive discussion of 
this result) and in a partial covering component with high covering fraction needed 
at the Y phase bin, which is not required at earlier or later phases, however (see Table~\ref{tab:doubleprs}).
A possible interpretation of these results is given by the oscillating behavior 
of the accretion wake \citep{Blondin+90}.
Because of the orbit-to-orbit variations of the ambient wind, the tail of the accretion wake 
is expected to oscillate with respect to our line of sight.
Averaging this behavior over several orbital periods would result in a large covering fraction,
where the wobbling accretion wake simulates an inhomogeneous absorbing environment.

A different interpretation might be that an intrinsically inhomogeneous structure 
is present around the inferior conjunction and partially covers the X-ray source.
Inhomogeneous stellar winds from supergiant stars (so-called clumps) are expected 
\citep{Runacres+Owocki05} and observed (\citealt{Goldstein+04}, and references therein), 
and have been suggested as the cause for \src~flaring activity 
(see \citealt{Kreykenbohm+08} and \citealt{Martinez+14}).
The timescale on which clumps hit the X-ray pulsar is on the order of the wind's dynamic timescale
\begin{equation}
\frac{0.5R_{\rm B}}{v_{\rm \infty}}\sim\frac{10^{11}\,cm}{1100\,km/s}\sim1000\,s
\end{equation}
where $0.5R_{\rm B}$ is the average separation between clumps \citep{Sundqvist+10}.
Thus, if we consider the X-ray source on longer timescales, the resulting effect is a partially covered spectrum.
However, if the partial covering is due to the inhomogeneous structure of the ambient wind,
no modulation of this component with orbital phase would occur.
In contrast, our results underline the necessity of partial covering only for the 
Y phase bin spectra, where the obscuration effect of the accretion wake is maximized.
Hydrodynamical simulations \citep{Blondin+90, Blondin+91} revealed that the accretion wake 
is composed of compressed filaments of gas and rarefied bubbles.
Moreover, recent models also showed that density perturbations in stellar winds of OB stars 
only develop at distances larger than the NS orbital radius in Vela X-1
($r\sim1.1\,R_{\star}$, see Fig.~$2$ in \citealt{Sundqvist+Owocki13}).
Therefore, intrinsic stellar wind clumps around the X-ray pulsar could be not 
structured enough to trigger the observed high X-ray flux variability.
Instead, the inhomogeneous structure characteristic of the accretion wake trailing the NS
indicates that the compact object could be able to perturb the ambient wind 
such that clumps are formed at its passage. 
The sphere of influence of the X-ray pulsar is dictated by its accretion radius, 
$r_{\rm acc}{\sim}10^{11}\,$cm \citep{Feldmeier+96} and by the Str{\"o}mgren sphere 
defined by the photoionization parameter $\xi=L_x / (n\,R_{\rm ion}^2)$, 
which leads (see Sect.~\ref{subsec:doublelc} for parameter definitions and values)
to a photoionization radius $R_{\rm ion}\sim0.2R_B\sim6\times10^{11}\,$cm\,$\sim r_{\rm acc}$.
Since the orbital velocity of the NS is $v_{\rm orbit}\approx280\,$km/s,
the perturbation timescale is about 
\begin{equation}
t_{\rm pert}=\frac{R_{\rm ion}}{v_{\rm orbit}}\approx2\times10^4\,s\sim5.5\,h
\end{equation}
which can be considered as an upper limit for the X-ray variability induced by accretion of locally formed clumps.
This is consistent with observations which showed that flares in \src~last from $10^2\,$s 
up to $10^4\,$s (see \citealt{Ducci+09, Martinez+14}, and references therein).

Even though the inhomogeneous structure characteristic of the accretion wake might be 
a possible cause for the observed partial covering around the inferior conjunction, 
it cannot be directly probed within the context of our \maxi~analysis, which only allows 
inferring the long-term properties of the source averaged over several orbital phase bins.

High-resolution spectroscopic observations carried out at critical orbital phases 
are necessary in future to distinguish among different models and constrain 
stellar wind scenarios and accretion modalities in \src.

\subsubsection{Standard sample phase-resolved spectroscopy}\label{subsec:standard_prs_results}

Modeling phase-resolved spectra of the standard sample with a 
cutoff power-law leads to a hardening of the spectra around the inferior conjunction 
of about $\Delta\Gamma{\sim}0.35$.
As we pointed out in Sect.~\ref{subsec:double_prs_results}, 
a physical mechanism for such an orbital dependence of the photon index is not known 
and is difficult to explain.

On the other hand, fitting the spectra with a partial covering model
shows that the neutral matter column density $N_{\rm H}$ 
increases toward later phases, varying from 
${\sim}10^{22}$ after the egress ($\phi_1$), to ${\sim}10^{23}\,$cm$^{-2}$ before the ingress ($\phi_7$).
This agrees with earlier studies of \src~(see, e.g., \citealt{Nagase89}),
and a similar phenomenon has been observed in other HMXBs as well, for instance, 
in 4U~$1700$-$37$ \citep{Haberl+89}.
Based on the analysis of both optical \citep{Carlberg78, Kaper+94} and 
X-ray data \citep{Feldmeier+96}, it has been shown that the accretion wake is not 
adequate to explain the later phase ($\phi>0.5$) absorption, and that a more extended 
photoionization wake is necessary.

When this model is applied, the spectral photon index modulation 
along the orbital phase is only marginal.
Based on the best-fit results, we can adopt here a similar interpretation 
as for the double-peaked sample results, that is, an inhomogeneous ambient medium 
that produces strong absorption of the X-ray emission and whose absorption efficiency 
depends on the orbital phase (see Sect.~\ref{subsec:double_prs_results}).
The column density shows a local peak of $N_{\rm H}^{pc}{\sim}3\times10^{23}\,$cm$^{-2}$ 
around the inferior conjunction, similar to what is observed for the double-peaked sample, 
strengthening the connection with the accretion wake.
Moreover, this sample needs the partial covering component also to fit the spectrum of the phase bin $\phi_2$,
which is located around the NS greatest eastern elongation (see Fig.~\ref{fig:sketch}).
Regardless of whether the partial covering is due to a wobbling or an inhomogeneous accretion wake,
these results indicate at a wake that affects the observations at earlier phases than the inferior conjunction.

We therefore conclude that the absorbing (and likely inhomogeneous) material 
that partially covers the X-ray pulsar might be due to a different 
influence of the accretion wake at different orbital phases.

\label{sec:discussion}

\section{Summary}
We have presented the results of a very long \maxi~observation of \src,
which allowed us to perform detailed spectral (phase-averaged and phase-resolved, in $2-20\,$keV) 
and light curve (in $4-20\,$keV) analysis. 
Our results can be summarized as follows:
\begin{itemize}

\item About $15\%$ of the orbital light curves in \src~shows a double-peaked orbital profile, 
      with a ${\sim}1.5\,$day dip around the inferior conjunction in the $4-10\,$keV and $10-20\,$keV energy bands.

\item Explaining the dip around the inferior conjunction by absorption alone requires column density values 
      of about $2\times10^{24}\,$cm$^{-2}$, which are not observed in our analysis.

\item The dip in the double-peaked sample might be produced by considering contribution from Thomson scattering 
      by an ionized accretion wake that is elongated toward the observer during the transit.
      Intrinsic variability of the stellar wind leads to strong variations of the accretion wake, including its 
      ionization degree, thus to alternating double-peaked and standard profiles.

\item When fit with a cutoff power-law model, the double-peaked and the standard samples both
      show spectral hardening around the inferior conjunction (by ${\sim}0.3$ in terms of the photon index).
      We did not find any reliable physical interpretation for this phenomenon, 
      which likely indicates that the model is inadequate for \src.

\item Orbital photon index modulation is avoided if a partial covering 
      component is used around the inferior conjunction.
      This component indicates a highly inhomogeneous environment and local column density 
      peak values of ${\sim}3\times10^{23}\,$cm$^{-2}$ around the inferior conjunction 
      (orbital phases around the eclipse were not analyzed).

\item Our results suggest either a wobbling or an inhomogeneous accretion wake 
      as the source of the partial covering around the inferior conjunction.
      Gravitational and radiative effects from the X-ray pulsar may be responsible of density inhomogeneities 
      in the ambient wind that successively feed the observed high X-ray variability.
      Moreover, the absorption properties of the accretion wake seem to take place 
      at earlier phases than the inferior conjunction.

\item Some of the analyzed orbital phase-bin spectra do not need a partial covering component.
      Therefore, any spectral analysis of \src~needs to consider the orbital phase at which the observation has been carried out.

\end{itemize}

\label{sec:conclusions}
\balance
\begin{acknowledgements}
We gratefully acknowledge the anonymous referee for numerous comments that greatly improved the manuscript.
This work is supported by the \textsl{International Program Associate} (IPA) of RIKEN, Japan, 
and by the \textsl{Bundesministerium f\"{u}r Wirtschaft und Technologie} 
through the \textsl{Deutsches Zentrum f\"{u}r Luft- und Raumfahrt e.V. (DLR)} under the grant number FKZ 50 OR 1204.
C.M. gratefully thanks the entire MAXI team for the collaboration and hospitality in RIKEN.
C.M. also thanks V. Doroshenko and L. Ducci (IAAT, T\"{u}bingen) for useful discussions.
\end{acknowledgements}


\bibliographystyle{aa} 
\bibliography{biblio}
\clearpage

\end{document}